\documentclass[prfluids,amsmath,amssymb,aps,floatfix]{revtex4-2}

\usepackage{hyperref}
\usepackage[utf8]{inputenc}
\usepackage{amsmath}
\usepackage{amsmath}
\usepackage[utf8]{inputenc}
\usepackage[T1]{fontenc}
\usepackage{graphicx}
\usepackage{float}
\usepackage{wrapfig}

\usepackage[usenames,dvipsnames]{xcolor}

\begin{document}

\title{Spontaneous vortex crystal formation in classical rotating flows} 

\author{Gabriel Marchetti and Pablo D. Mininni}
\affiliation{Universidad de Buenos Aires, Facultad de Ciencias Exactas y Naturales, Departamento de Física, Ciudad Universitaria, 1428 Buenos Aires, Argentina,}
\affiliation{CONICET - Universidad de Buenos Aires, Instituto de F\'{\i}sica Interdisciplinaria y Aplicada (INFINA), Ciudad Universitaria, 1428 Buenos Aires, Argentina.}

\date{\today}

\begin{abstract}
    Vortex crystals, ordered structures observed in superconductors and rotating superfluids, have also been hypothesized to form in classical fluids, based on numerical simulations and observations of the Jovian polar atmospheres. We perform direct numerical simulations of the Navier–Stokes equations in rotating frames, to investigate the spontaneous emergence of metastable vortex crystals. We analyze the energy spectrum, vortex morphology, and spatio-temporal dynamics to understand their roles in crystal formation and evolution. In addition, we explore domains with varying aspect ratios to examine their impact on the vortex lattice. Our results indicate a relationship between the crystal lifespan and dissipation, and we propose a scaling law linking the rotation rate, domain geometry, and vortex lattice periodicity. Finally, we identify a critical threshold in the control parameter, the Rossby number, suggesting a behavior similar to that found in phase transitions.
\end{abstract}

\maketitle

\section{Introduction}

Nonlinearity is one of the defining features of fluid dynamics, especially in the turbulent flow regime. Turbulence encompasses a myriad of interacting structures spanning a broad range of spatial and temporal scales. This property leads to disorder and chaotic behavior that is unrepeatable from one experiment to the next, yet with repeatable statistics. However, it is not only disorder that governs turbulent flows, but also order. Self-organization mechanisms in turbulence have been a subject of study for several decades, and in recent years have been observed in a variety of configurations \cite{kraichnan_2d, driscoll1, driscoll2, jimenez, Xia_2011, Sen_2012, jupiter, Pouquet_2017, Alexakis_2024}. In this process, the chaotic motion of small eddies results in spontaneous macroscopic order. Canonical examples of this phenomenon include the inverse energy cascade and the energy condensate in two-dimensional flows \cite{kraichnan_2d, Boffetta_2012} and in space plasmas \cite{hossain, Mininni_2005, Alexakis_2006, Mininni_2007}. The spontaneous appearance of order has recently been understood as analogous to phase transitions in thermodynamic systems, governed by a system control parameter \cite{biferale, ABB2018}. However, with the exception of condensates in ideal flows \cite{kraichnan_2d}, the self-organization in forced and viscous cases take place in conditions that are far from equilibrium and where traditional thermodynamic ideas do not apply.

Rotating flows provide an interesting example of inverse energy cascades and self-organization. The theory of rotating turbulence has been developed in detail \cite{wallefe93, cambon, Cambon_2004}, and in the limit of infinite rotation and for infinite domains, two-dimensional modes are decoupled from three-dimensional modes which transfer their energy solely to smaller scales. Ideal truncated rotating flows also lack a condensate in their statistical equilibrium, displaying only thermalization of the energy, and disorder \cite{Mininni_2011}. However, for finite rotation and in finite domains inverse energy cascades develop \cite{smith99, Sen_2012, campagne, buzzicotti}. This results in the formation of strong columnar vortices. While the stability of columnar structures in rotating flows can be understood in terms of the Proudman--Taylor theorem \cite{davidsonrot}, it is the nonlinearity that is responsible for their formation in the dynamical case. As a result, energy injected at intermediate scales is transferred both to large and small scales, with some energy leaking from three-dimensional modes to two-dimensional modes, resulting in a partial flow bi-dimensionalization \cite{wallefe93, smith99, Sen_2012, ABB2018}. These picture has been confirmed in experiments \cite{campagne}, which showed that while rapidly rotating turbulence in finite domains results mainly in an inverse cascade of energy to large scales, at intermediate rotation strengths a split cascade develops in which energy is transferred in both directions, and eventually for sufficiently low rotation energy only goes to smaller scales.

A remarkable example of organized states in this context is given by vortex crystals. Recently, self-organized metastable states of vortex crystals in rotating flows were observed in \cite{clark}. The parameters that control the lattice formation, size, and stability are unclear, as the inverse energy cascade in rotating turbulence is expected to always proceed to the largest available scale instead of being arrested at some intermediate scale. But the resulting vortex lattice is reminiscent of other vortex crystals observed in condensed matter. Perhaps the most typical example of such vortex crystals is given by Bose-Einstein condensates \cite{newton, Fetter_2009}, particularly in magnetic vortices in superconductors \cite{Brandt_1986}, and in vortices in rotating superfluids and gaseous condensates \cite{Fetter_2009, abo, Peretti_2023, amette}. These crystals organize in triangular lattices, commonly referred to as Abrikosov lattices. Vortex crystals have also been observed in some classical flows. Beside the case of classical rotating flows described in \cite{clark}, examples are provided by electron systems subjected to a magnetic field with a large relaxation time \cite{driscoll1, driscoll2}, two-dimensional turbulent flows in which a balance of deterministic and random forcing was used to sustain a stable vortex crystal in a background vorticity \cite{jimenez}, and chiral liquid crystals \cite{Fernandez_2024}. Recently, vortex crystals were also reported in fluids with odd viscosity \cite{dewit_2024}. And observations of stable vortex arrays were reported by NASA Juno probe in the Jovian polar atmospheres \cite{jupiter}. This polar structures feature vortices with cyclonic behavior arranged in rotationally symmetric patterns, with the north pole featuring eight cyclones orbiting a central one, while the north pole features five cyclones.

In this work, we study the spontaneous formation of vortex crystals in rotating flows. Our main objective is to identify under what conditions flows develop ordered vortex crystals by varying the physical parameters of the system, and to identify the parameters that control the lattice spatial periodicity and stability. To this end, we perform direct numerical simulations (DNSs) of the Navier-Stokes equations in a rotating frame. We consider the main three regimes observed in rotating flows, including cases with an inverse cascade of energy, a direct cascade of energy, and a flux-loop regime \cite{clark} in which the inverse cascaded energy returns to smaller scales resulting in accumulation of energy at intermediate scales. Normal viscosity is used, which results in DNSs with moderate Reynolds numbers, but allows us to quantify the role of physical viscosity in the lattice stability. We consider the time evolution of the flow kinetic energy, as well as its spectrum, to identify regimes in which the inverse energy cascade stalls at an intermediate scale between the forcing scale and the largest scale in the system. These regimes are metastable, and eventually energy continues its transfer towards larger scales and is also partially dissipated. During the metastable regime, we study the lattice geometry as a function of the controlling parameters. The analysis allows us to identify a critical threshold in the Rossby number to have vortex crystals, a dependence of the lattice periodicity with the Rossby number, and a linear dependence of the time of stability of the crystal with the viscous time.

\section{Methodology}

We use DNSs of the incompressible velocity field $\bf u$ described by the Navier-Stokes equation in a rotating frame,
\begin{equation}
    \frac{\partial \mathbf{u}}{\partial t} + \mathbf{u}\cdot \boldsymbol{\nabla}\mathbf{u} = -\boldsymbol{\nabla} P - 2\mathbf{\Omega}\times\mathbf{u} + \nu\nabla^2\mathbf{u} + \mathbf{f},
\end{equation}
where $\boldsymbol{\nabla} \cdot \mathbf{u} = 0$, $P=p/\rho - \Omega^2r^2/2$ is the fluid pressure $p$ divided by the uniform mass density $\rho$, corrected  by the centrifugal pressure per mass unit $-\Omega^2r^2/2$, the rotation angular velocity is ${\bf \Omega} = \Omega \hat{z}$ with $\Omega$ the rotation frequency, and $\nu$ is the kinematic viscosity. Simulations were perfomed in Cartesian, triply periodic domains of dimension $2\pi L_x \times 2\pi L_y \times 2\pi L_z$, where all lengths are measured in units of a unit length $L_0$. The DNSs were done using the parallel pseudospectral code GHOST \cite{ghost, Rosenberg_2020}. Energy was injected into the flow at rate $\epsilon$, using the mechanical forcing $\bf f$ composed of a superposition of harmonic modes acting in the wave number window $[k_f,k_f+2]$, where $k_f$ is the forcing wave number.

The dimensionless control parameters of the system are the Rossby number at the forcing scale, $\mathrm{Ro}_f$, and the Reynolds number at the forcing scale, $\mathrm{Re}_f$, respectively given by
\begin{align}
    \mathrm{Ro}_f= \frac{(\epsilon k_f^2)^{1/3}}{\Omega},&&
    \mathrm{Re}_f = \frac{(\epsilon k_f^{-4})^{1/3}}{\nu},
\end{align}
and the parallel and perpendicular aspect ratios of the domain, defined respectively as
\begin{align}
    \lambda_\parallel = \frac{L_z}{\sqrt{L_xL_y}},&&
    \lambda_\perp = \frac{L_y}{L_x}.
\end{align}
Based on these definitions, we define the turnover time at the forcing scale as $\tau_f = (\epsilon k_f^2)^{-1/3}$. 

\begin{table}
    \centering
    \begin{ruledtabular}
    \begin{tabular}{cccccccccccc}
        \multicolumn{1}{c}{Simulation} &
        \multicolumn{1}{c}{$N_x$} &
        \multicolumn{1}{c}{$N_y$} &
        \multicolumn{1}{c}{$N_z$} &
        \multicolumn{1}{c}{Aspect ratio} &
        \multicolumn{1}{c}{$\nu T_0/L_0^2$} &
        \multicolumn{1}{c}{$\Omega T_0$} &
        \multicolumn{1}{c}{$k_f L_0$} &
        \multicolumn{1}{c}{$\epsilon L_0/U_0^3$} &
        \multicolumn{1}{c}{$N_V$} &
        \multicolumn{1}{c}{$\mathrm{Ro}_f$} \\
        \hline
        I & 256 & 256 & 256 & 1 : 1 : 1 & $1.5\times10^{-3}$ & 3 & 10 & 0.19 & 0 & 0.89 \\
        II & 256 & 256 & 256 & 1 : 1 : 1 & $1.5\times10^{-3}$ & 4 & 10 & 0.18 & 1 & 0.66 \\
        III & 256 & 256 & 256 & 1 : 1 : 1 & $1.5\times10^{-3}$ & 4.5 & 10 & 0.17 & 2 & 0.58 \\
        IV & 384 & 384 & 384 & 1 : 1 : 1 & $8.6\times10^{-4}$ & 5.5 & 15 & 0.17 & 5 & 0.62 \\
        V & 384 & 384 & 384 & 1 : 1 : 1 & $8.6\times10^{-4}$ & 6 & 15 & 0.17 & 6 & 0.56 \\
        VI & 512 & 512 & 512 & 1 : 1 : 1 & $5.9\times10^{-4}$ & 6.5 & 20 & 0.16 & 6 & 0.62 \\
        VII & 512 & 512 & 512 & 1 : 1 : 1 & $5.9\times10^{-4}$ & 7 & 20 & 0.16 & 7 & 0.57 \\
        VIII & 512 & 512 & 512 & 1 : 1 : 1 & $5.9\times10^{-4}$ & 7.5 & 20 & 0.16 & 8 & 0.53 \\
        IX & 256 & 432 & 256 & 1 : $\sqrt{3}$ : 1 & $1.5\times10^{-3}$ & 4.5 & 10 & 0.17 & 2 & 0.58 \\
        X & 512 & 864 & 512 & 1 : $\sqrt{3}$ : 1 & $5.9\times10^{-4}$ & 7 & 20 & 0.16 & 12 & 0.57 \\
        XI & 512 & 432 & 512 & 1 : $\sqrt{3}/2$ : 1 & $5.9\times10^{-4}$ & 7 & 20 & 0.17 & 6 & 0.58 \\
        XII & 256 & 256 & 512 & 1 : 1 : 2 & $1.5\times10^{-3}$ & 4.5 & 10 & 0.18 & 4 & 0.58 \\
        XIII & 384 & 384 & 768 & 1 : 1 : 2 & $8.6\times10^{-4}$ & 6 & 15 & 0.17 & 4 & 0.56 \\
        XIV & 384 & 384 & 768 & 1 : 1 : 2 & $8.6\times10^{-4}$ & 6.5 & 15 & 0.17 & 5 & 0.52 \\
        XV & 384 & 384 & 768 & 1 : 1 : 2 & $8.6\times10^{-4}$ & 10 & 15 & 0.57 & 7 & 0.50 \\
        XVI & 384 & 384 & 768 & 1 : 1 : 2 & $8.6\times10^{-4}$ & 15 & 15 & 1.5 & 7 & 0.46 \\
        \noalign{\vskip .7mm}
    \end{tabular}
    \end{ruledtabular}
    \caption{Numerical simulations and their parameters. Resolutions $N_x$, $N_y$, and $N_z$, in each direction, correspond to the linear number of grid points. The aspect ratio gives the length of the domains in the $x$, $y$, and $z$ directions in units of $2 \pi L_0$. The kinematic viscosity is $\nu$, $\Omega$ is the angular velocity of rotation, $k_f$ the forcing wavenumber, $\epsilon$ the energy injection rate, $N_V$ is the number of vortices in the crystal formed, and $\mathrm{Ro}_f$ is the Rossby number at the forcing scale. Dimensional quantities are made dimensionless using the unit length $L_0$ and the unit time $T_0 = L_0/U_0$, where $U_0$ is the unit velocity.}
    \label{tab:sims}
\end{table}

To obtain different vortex crystals we performed simulations varying the dimensional control parameters of the system, using the following protocol. We maintained a fixed forcing amplitude and varied $\Omega$, to probe different values of $\mathrm{Ro}_f$. The Rossby number $\mathrm{Ro}_f$ was varied to explore moderate values of rotation, in order to focus on the transition between flows with direct and inverse energy cascades. The forcing wave number and the viscosity were varied keeping $k_f$ of $\mathcal{O}(k_\eta/10)$, where $k_\eta = (\epsilon/\nu^3)^{1/4}$ is the Kolmogorov dissipation wave number, to study the effect of $k_f$ and of the viscous dissipation time $\tau_\eta = (\nu/\epsilon)^{1/2}$ on the vortex crystal stability. The spatial resolution was changed to keep $k_\eta < k_\textrm{max}$, where $k_\textrm{max}$ is the maximum resolved wave number in the simulation. Finally, the aspect ratios of the domain $\lambda_\parallel$ and $\lambda_\perp$ were varied to study the effect of the domain geometry in the formation and the characteristic scale of the crystals. Most DNSs are listed in table \ref{tab:sims}, in which dimensional quantities are written in units of the unit length $L_0$, a unit velocity $U_0$, and a unit time scale $T_0 = L_0/U_0$ (a few more simulations, not listed in the table, were performed to use as reference cases and will be briefly discussed below). Rewriting $\mathrm{Re}_f$ in terms of $k_\eta/k_f$ yields $\mathrm{Re}_f=(k_\eta/k_f)^{4/3}$, meaning that in all our simulations $Re_f\approx 21$.

\section{Results}

\subsection{Energy evolution and energy spectra}

\begin{figure}
    \centering
    \includegraphics[width=0.85\linewidth]{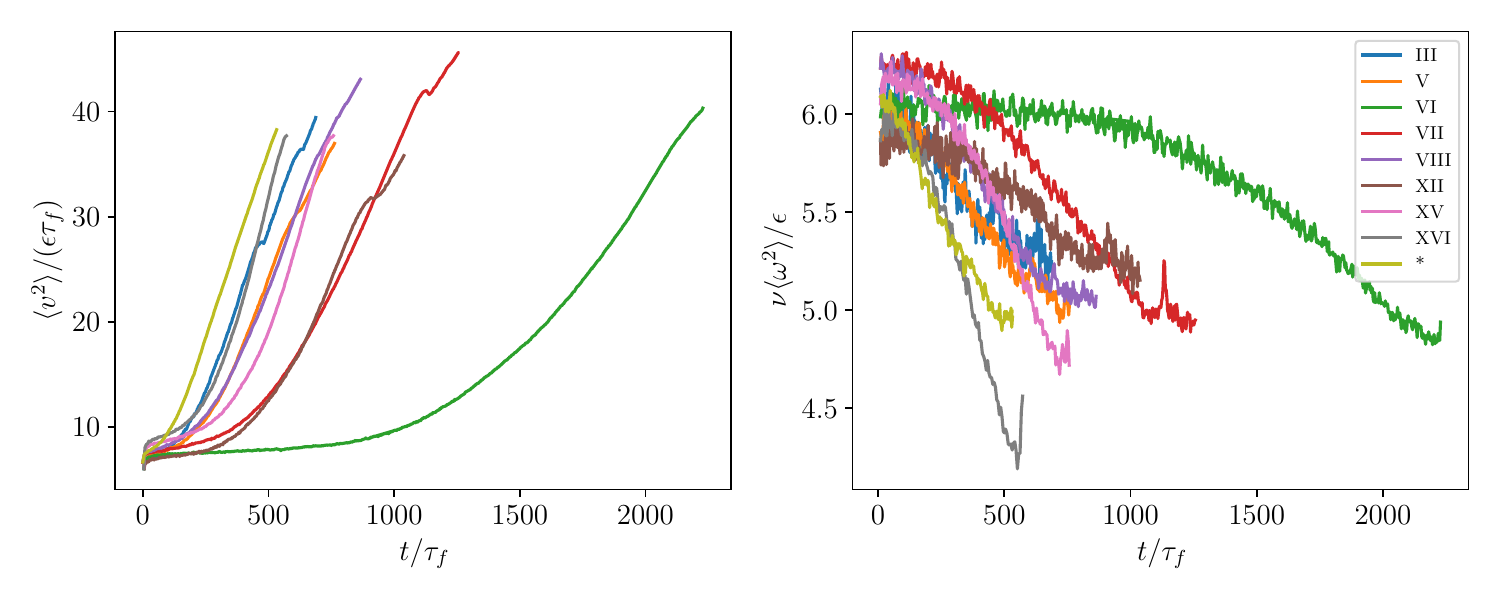}
    \caption{Time evolution of $\langle v^2\rangle$, proportional to the kinetic energy in the system and normalized by $\epsilon \tau_f$ ({\it left}), and of the energy dissipation rate $\nu \langle \omega^2\rangle$ normalized by $\epsilon$ ({\it right}), for several simulations in table \ref{tab:sims} (see the inset for the labels). The starred curve shows as a reference a simulation with a typical inverse energy cascade. Note that most of the other simulations display a dip in the growth of $\langle v^2\rangle$ at some moment, accompanied by a local peak in the energy dissipation rate.}
    \label{fig:energy}
\end{figure}

We first study the total energy and its spectral evolution in the simulations in table \ref{tab:sims}. Figure \ref{fig:energy} shows $\langle v^2\rangle$ normalized by $\epsilon \tau_f$, as a function of time in units of $\tau_f$. The time evolution of the energy dissipation rate, $\nu \langle \omega^2\rangle$, normalized by $\epsilon$, is also shown for the same simulations. In addition to several simulations from table \ref{tab:sims} that develop a metastable vortex crystal, in Fig.~\ref{fig:energy} we also show as a reference the time evolution of a simulation with a Rossby number small enough that the system displays a strong and steady inverse energy cascade without developing a transient crystal. In all the other simulations the kinetic energy grows until, at a certain moment, the energy and the dissipation rate evolution change their behavior. The energy shows a dip before growing again, and at the same time the energy dissipation rate displays a pronounced local peak. As will be seen later, this time corresponds to the instability of a vortex crystal metastable state.

\begin{figure}
    \centering
    \includegraphics[width=0.85\linewidth]{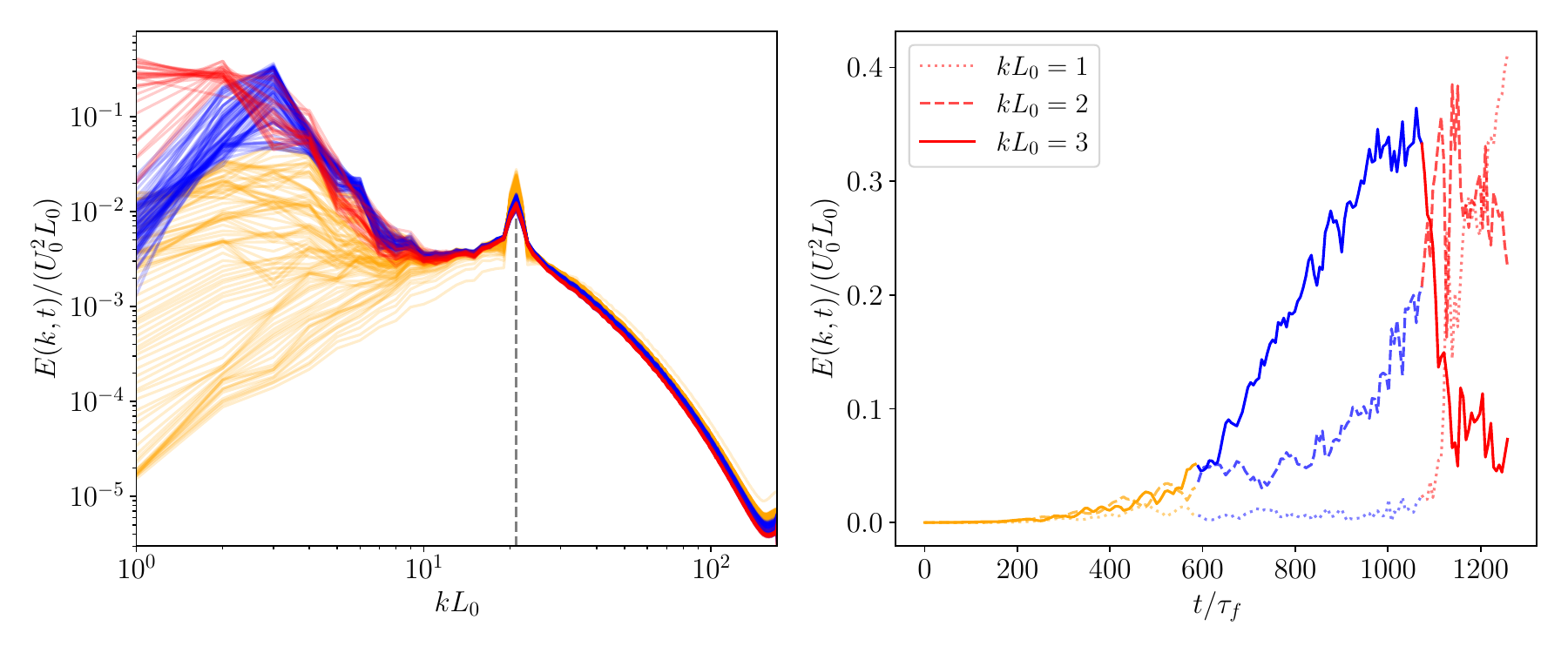}
    \caption{{\it Left:} Time evolution of energy spectrum in simulation VII. Energy flowing from the injection scale (indicated with a dashed line) to larger scales stalls at an intermediate scale $l^*=2\pi/k^*$ (associated to the periodicity of the vortex crystal in the system), resulting in a peak at $k^*$. The spectra at times when energy peaks around $k^*$, and a vortex crystal is visible, is marked in blue. Orange and red lines indicate spectra at previous and later times, respectively. {\it Right:} Time evolution of the energy in the Fourier shells $k L_0=1$, 2, and 3 in the same simulation.
    }
    \label{fig:spectra}
\end{figure}

\begin{figure}
    \centering
    \includegraphics[width=.65\linewidth]{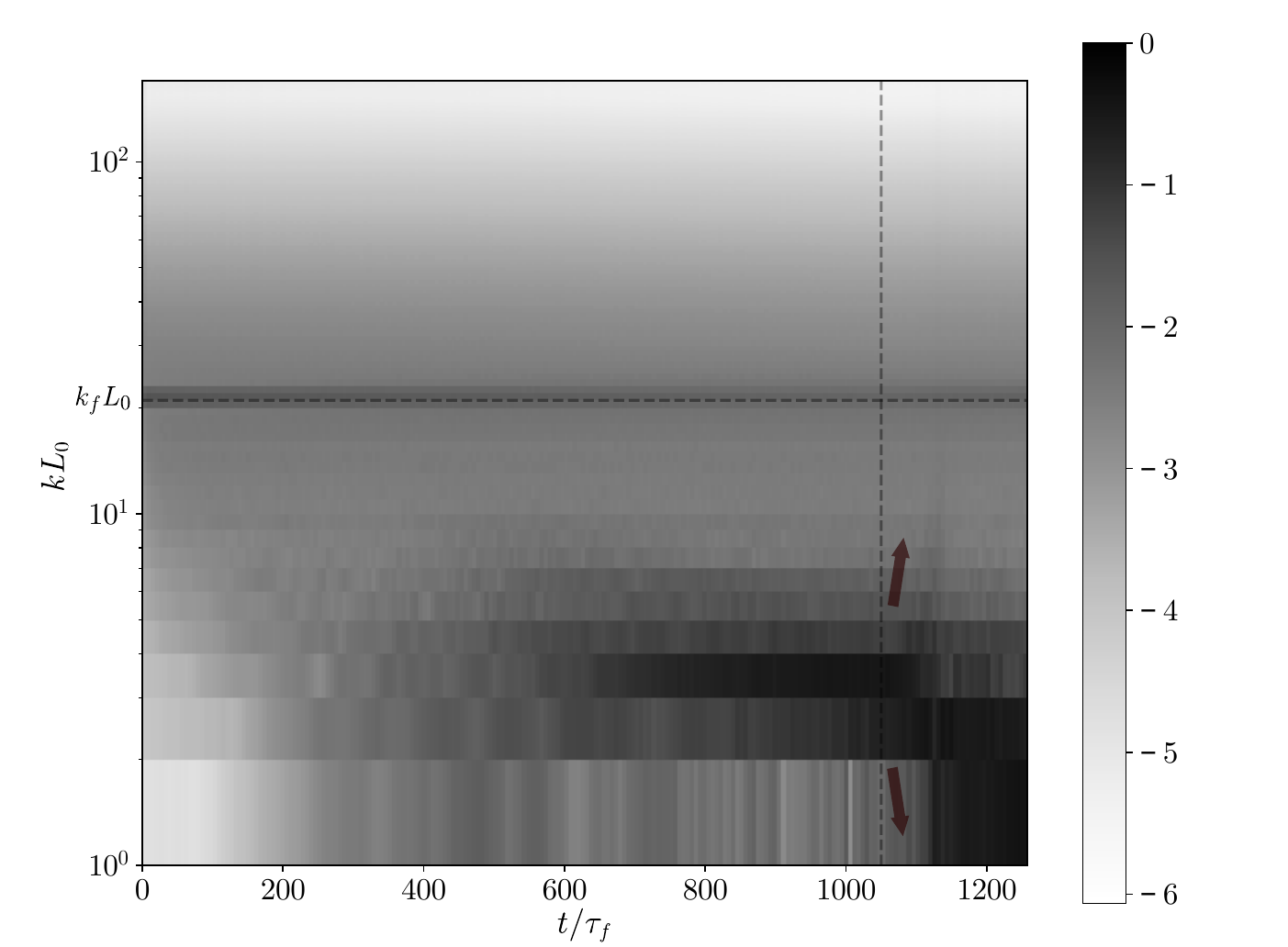}
    \caption{Spectrogram of simulation VII; the gray scale indicates $\log (E(k)/(U_0^2 L_0))$. Between $t / \tau_f \approx 600$ and $1050$, energy that inverse cascades from $k_f$ accumulates at an intermediate wave number $k^*$ (see the dark band around $kL_0 = 3$). At a time $t/\tau_f\approx 1050$ energy is redirected mostly to the smallest available wave number (lower arrow), with a small fraction of the energy propagating towards larger wave numbers (upper arrow).}
    \label{fig:spectrogram}
\end{figure}

Figure \ref{fig:spectra} shows the time evolution of the energy spectrum in simulation VII, and the detailed time evolution of the energy in the first three Fourier shells ($kL_0=1$, 2, and 3). The spectrum has a peak at $k_f$, and as time advances, energy grows at larger scales (i.e., at wave numbers smaller than $k_f$). However, at intermediate times, a clear peak develops at $k^* L_0=3$ (where we define $k^* = 2\pi/\ell^*$ as the wave number of the peak, and $\ell^*$ as its associated length scale), and the inverse transfer of energy does not proceed further towards smaller wave numbers. Instead, energy piles up at $k^*$ and the amplitude of its peak keeps increasing, until eventually the process stops and energy is abruptly transferred towards $kL_0=2$ and 1, with the energy at $k^* L_0=3$ suddenly decreasing. This is more clearly seen in the left panel of Fig.~\ref{fig:spectra}. At a time around $t/\tau_f \approx 1050$ the energy stalling mechanism that accumulated energy at $k^* L_0=3$ ends, and the energy at $kL_0=1$ grows rapidly. This is the same time seen in Fig.~\ref{fig:energy} for this simulation, at which the total kinetic energy displayed a dip, and the instantaneous energy dissipation rate displayed a peak. This indicates that the energy accumulated for a while at $k^* L_0=3$, is later partially transferred to larger scales and partially dissipated. Indeed, a small fraction of the energy at $k^* L_0=3$ is transferred to smaller scales (or larger wave numbers) where it can dissipate, as shown in the spectrogram in Fig.~\ref{fig:spectrogram}. Note how after $t/\tau_f \approx 1050$, an excess of spectral energy density can be seen propagating from $kL_0=3$ towards larger values of $k$. This combined direct and inverse energy transfer can be also confirmed by studying the energy transfer and flux (not shown). After this change in the flow behavior, the inverse cascade proceeds normally and energy piles up at $kL_0=1$.

\subsection{Flow visualization}

\begin{figure}
    \centering
    \includegraphics[width=.9\linewidth]{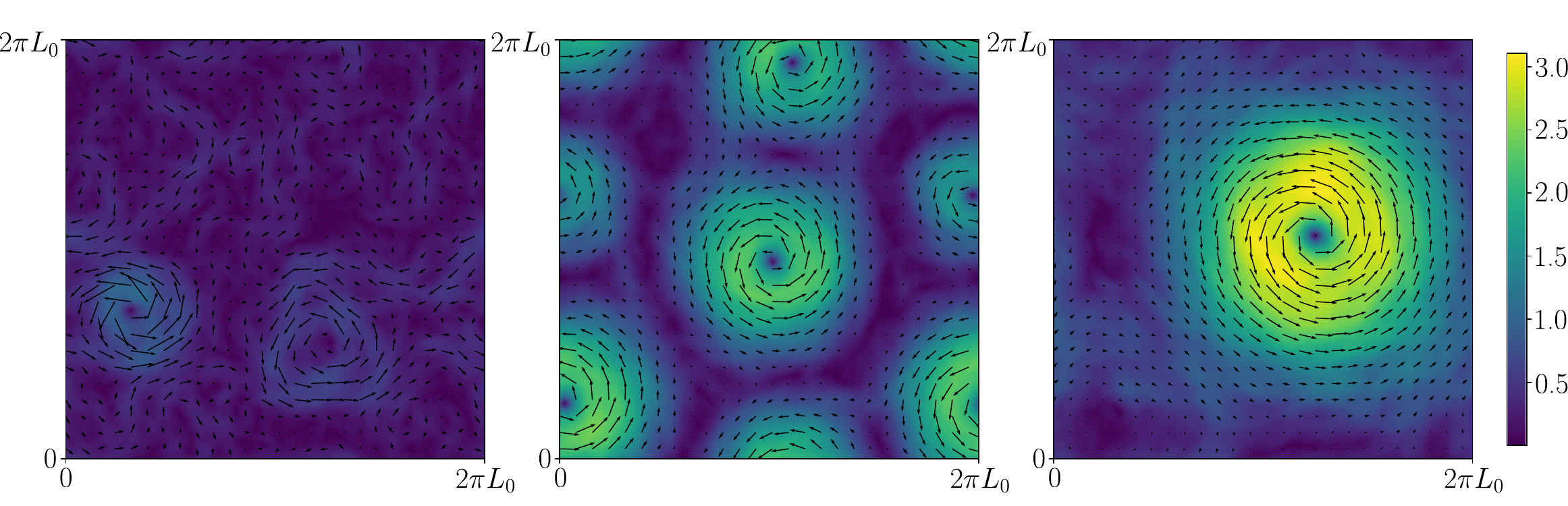}
    \caption{Velocity in the plane perpendicular to rotation, averaged along the rotation direction, in simulation XII. {\it Left:} Early times, at which a few large-scale vortices emerge from a disordered background. {\it Middle:} Intermediate times; note the formation of a triangular lattice with 4 vortices considering the periodic boundary conditions. {\it Right:} Late times, after the lattice became unstable, with all the vortices coalescing into one large-scale structure.
    }
    \label{fig:evolution}
\end{figure}

The transient phase observed in Figs.~\ref{fig:spectra} and \ref{fig:spectrogram} corresponds to a vortex crystal. In this section we confirm this by direct visualization of these metastable structures, with special emphasis on the spatial distribution of the vorticity and the velocity field. Figure \ref{fig:evolution} shows the flow at three different times in simulation XII, spatially averaged along the direction of rotation. The color indicates the magnitude of the averaged velocity components perpendicular to rotation, $v_\perp = |\langle \mathbf v - (\mathbf v\cdot\mathbf{\hat\Omega}) \mathbf{\hat\Omega}\rangle_\parallel|$, and the arrows indicate their directions, at three different times. In the left panel, corresponding to early times, large-scale vortices start to form from a disordered flow, fed by the inverse energy cascade. Later, in the mid panel, a fixed number of vortices organize in a lattice and accumulate energy. This structure survives for a long time. Finally, as shown in the right panel, the lattice becomes unstable and the inverse cascade continues, resulting in a large-scale vortex in the entire domain.

\begin{figure}
    \centering
    \includegraphics[width=.9\linewidth]{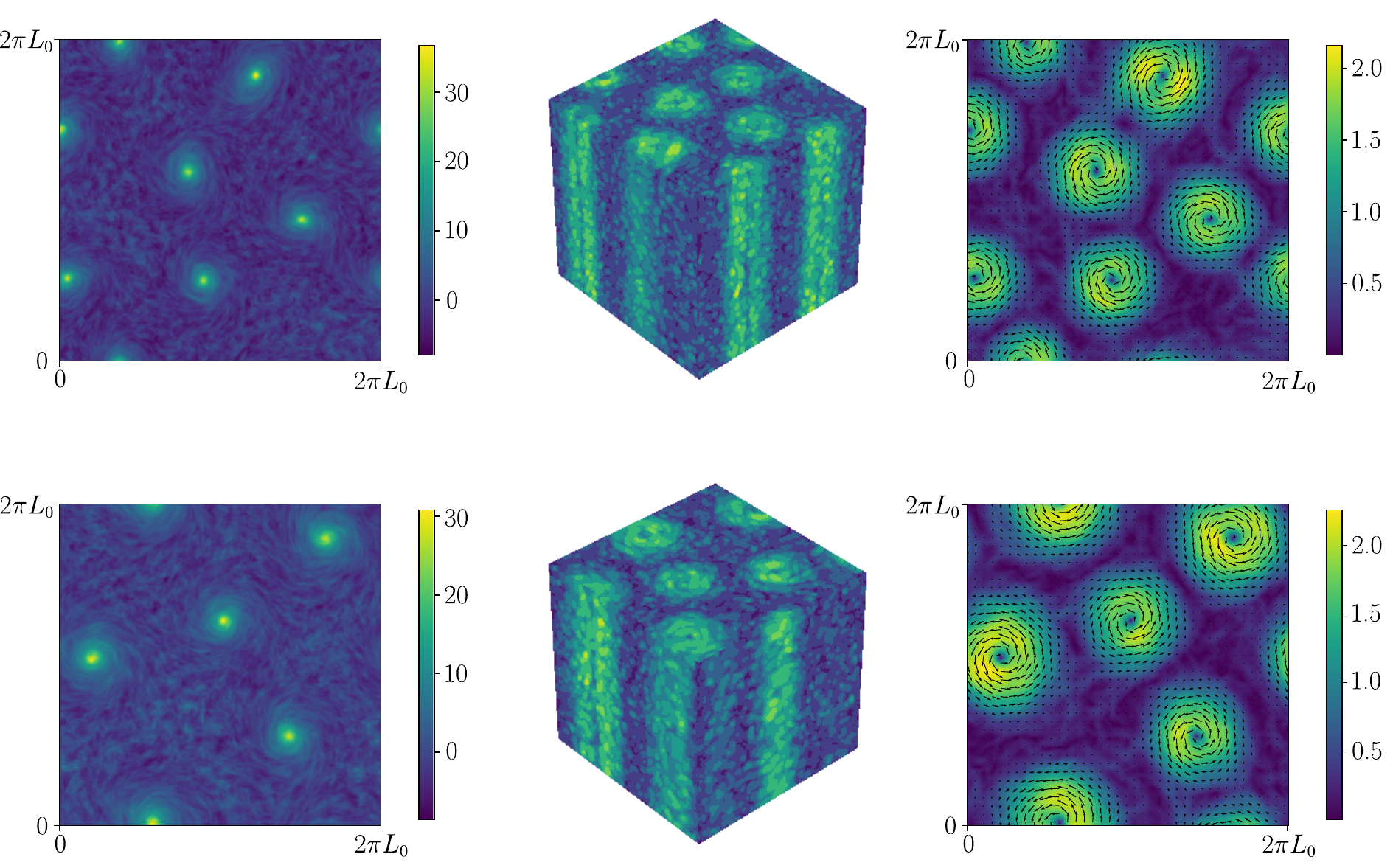}
    \caption{Vertically averaged vertical vorticity $\omega_\parallel$ ({\it left}), three-dimensional visualizaton of the perpendicular velocity intensity ({\it center}), and vertically averaged perpendicular velocity ({\it right}), in simulations VII ({\it top}) and IV ({\it bottom}). The flux–loop mechanism stalls the energy transfer at intermediate scales, resulting in lattices with defects resembling Abrikosov lattices in rotating superfluids. Note that all vortices in the lattices are cyclonic.}
    \label{fig:visualization}
\end{figure}

Figure \ref{fig:visualization} displays the vorticity component parallel to the rotation, spatially averaged along the same direction, $\omega_\parallel = \langle \boldsymbol\omega \cdot \mathbf{\hat\Omega} \rangle_\parallel$, as well as the previously defined perpendicular velocity $v_\perp$, for simulations VII and IV. A three-dimensional visualization of the perpendicular velocity is also shown to confirm the statistical translational symmetry of the lattice along the rotation direction. In all cases the lattices are triangular, but the number of vortices in the lattice depends on the parameters, with simulation VII displaying 7 vortices, and simulation IV displaying 5 vortices. Exploration of table \ref{tab:sims} indicates that this number is not solely controlled by the forcing scale, as simulations with the same value of $k_f$ can have different number of vortices.

Indeed, these lattices have a characteristic length larger than $2\pi/k_f$, and are generated by a self-organization process arrested by the flux-loop mechanism as already shown in \cite{clark}, instead of being caused directly by the forcing acting at $k_f$. Moreover, the periodicity of these lattices corresponds to the inverse wave number at which the energy spectrum peaks (see, e.g., Fig.~\ref{fig:spectra}). The periodicity seems to depend as a result not only on $k_f$ but also on $\Omega$, or in terms of dimensionless control parameters, on $\mathrm{Ro}_f$ as will be discussed in more detail in Sec.~\ref{sec:control}.

Finally, the visualizations in Figs.~\ref{fig:evolution} and \ref{fig:visualization} indicate that all vortices in the lattice are cyclonic. This is a manifestation of the cyclone-anticyclone asymmetry already reported in rotating turbulence \cite{smith2005, moisy2010, gallet2014}. Strong positive vorticity $\omega_\parallel$ accumulates in the center of these vortices. But as a result of the boundary conditions, the distribution of $\omega_\parallel$ must satisfy the constraint that the total circulation over the domain in the rotating frame must be zero. Therefore, in the region between the vortices $\omega_\parallel$ must have the opposite sign. This background of opposite-sign vorticity is approximately homogeneous, a feature also reported experimentally in other systems with vortex crystals \cite{driscoll2}.

\begin{figure}
    \centering
    \includegraphics[width=.9\linewidth]{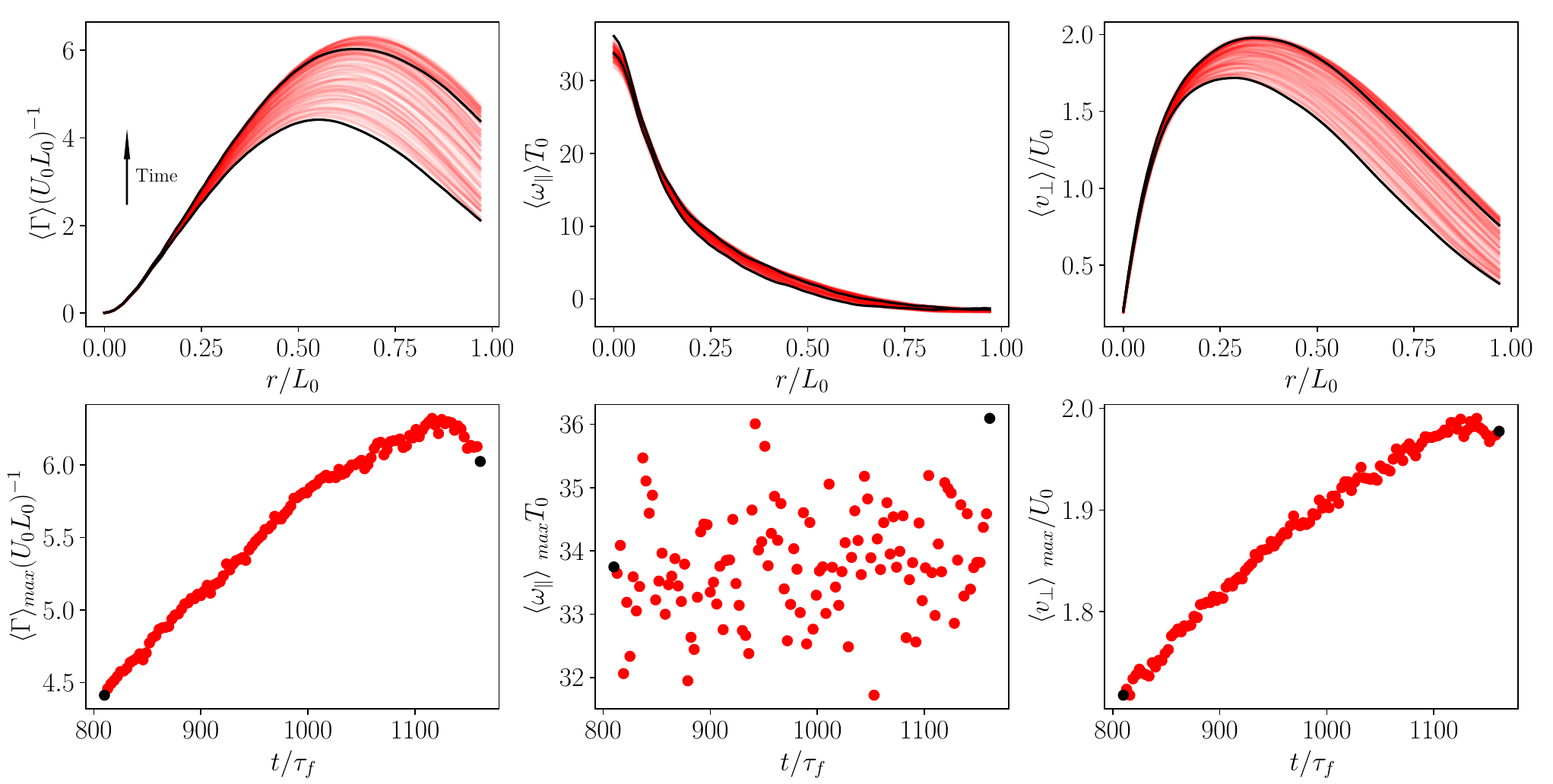}
    \caption{{\it Top:} Mean radial profiles of circulation ({\it left}), parallel vorticity ({\it middle}), and perpendicular velocity ({\it right}) of individual vortices in the crystal in simulation II. The time evolution of the mean profiles is shown, from the beginning of the crystal to its end (marked by the black lines). The arrow indicates the direction of time. {\it Bottom:} Time evolution of the maximum of circulation ({\it left}), parallel vorticity ({\it middle}), and perpendicular velocity ({\it right}) in the same simulation. The black dots correspond again to the initial time of crystal, and the final time close to the moment of the crystal's destruction.}
    \label{fig:circ}
\end{figure}

\begin{figure}
    \centering
    \includegraphics[width=0.4\linewidth]{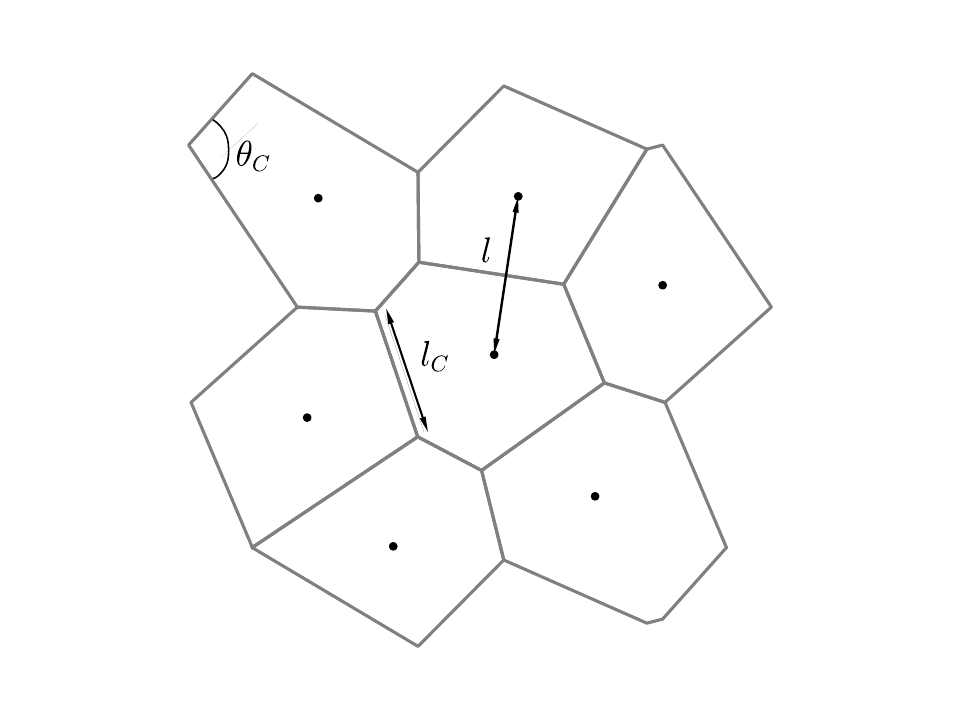}
    \caption{Example of a Voronöi tesselation using the vortex lattice in simulation VII. We define the inter-vortex length $l$ as the distance between centers, the cell side $l_C$ as the distance between vertices, and the angles $\theta_C$ as the cell inner angles.}
    \label{fig:voro}
\end{figure}

\begin{figure}
    \centering
    \includegraphics[width=.9\linewidth]{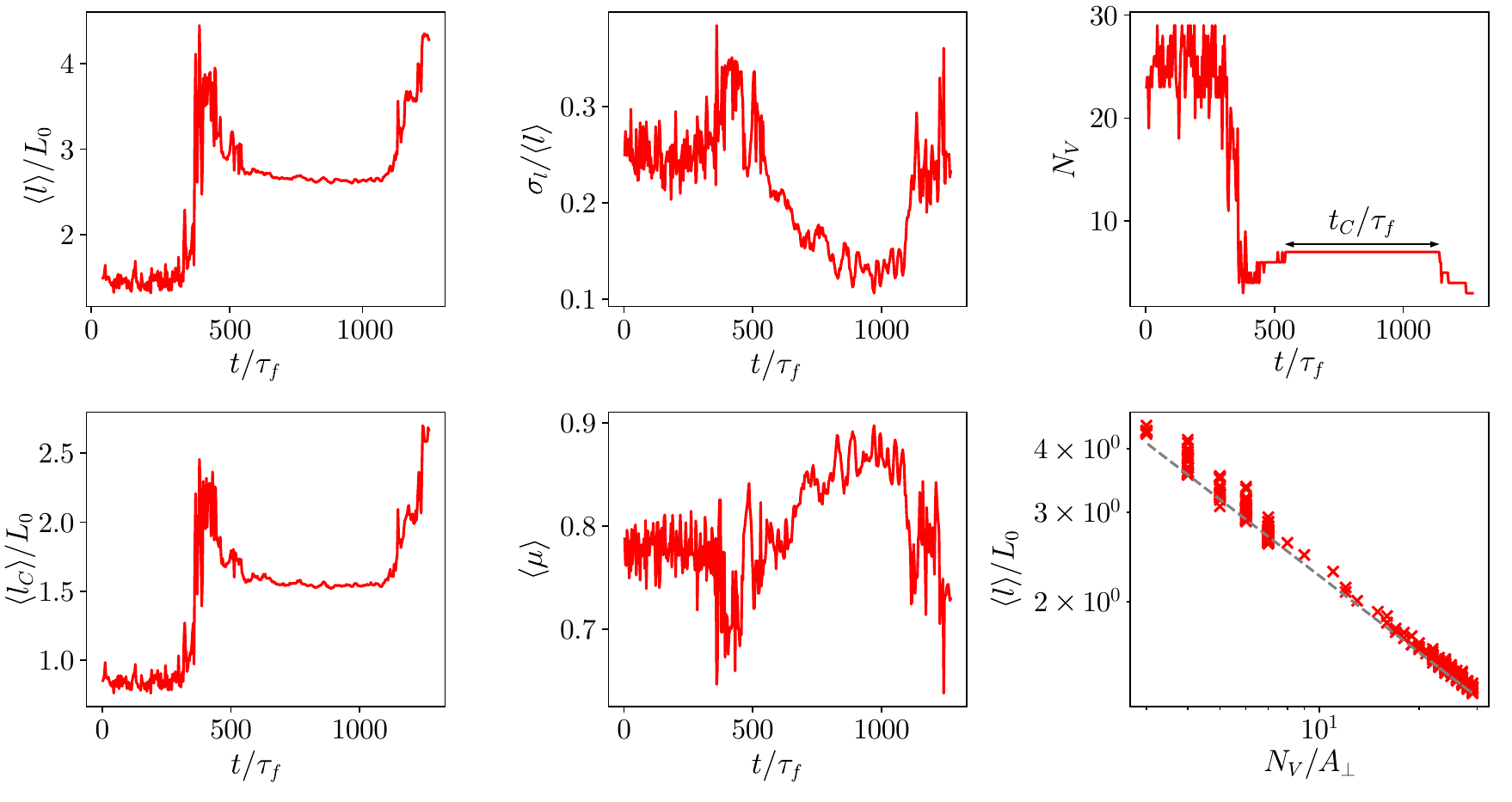}
    \caption{From left to right, and from top to bottom: Time evolution of the inter-vortex mean length $\langle l \rangle$, of the relative dispersion of the inter-vortex mean length $\sigma_l /\langle l \rangle$, of the number of vortices $N_V$, of the mean length of the Voronöi cell sides $\langle l_C\rangle$, and cell regularity $\mu$, for simulation VII. The crystal time is defined as the time for which $N_V$ remains constant, and is indicated in the top right panel. The bottom right panel shows $\langle l \rangle/L_0$ as a function of the density of vortices $N_V/A_\perp$ (where $A_\perp$ is the horizontal area of the domain) for the same simulation. The dashed line is the theoretical prediction in Eq.~(\ref{eq:voronoi}).}
    \label{fig:stats}
\end{figure}

\subsection{Spatio-temporal characterization \label{sec:spatiotemp}}

In order to identify the relevant parameters in the system, we have to properly define the time in which the crystal remains quasi-stable, and characterize the stages of evolution of the crystal and of the individual vortices in more detail, describing how different metrics evolve during crystal formation and decay. In this section we study the evolution of several flow magnitudes for individual vortices, and we introduce metrics to characterize the lattice.

We identify individual vortices by extracting local maxima of parallel vorticity $\omega_\parallel$. Once the centers were identified, we compute radial profiles of the angular and vertically averaged circulation $\Gamma$, parallel vorticity $\omega_\parallel$, and perpendicular velocity $v_\perp$. The results for simulation II are shown in Fig.~\ref{fig:circ}. Radial profiles are only shown for the times in which the lattice is present. The circulation grows in time (both the radial profile as well as its maximum). The same happens with the perpendicular velocity. However, at a certain moment the circulation and the perpendicular velocity reach a maximum, and then decrease. This is the time of the dip in the total kinetic energy and local maximum of dissipation observed in Fig.~\ref{fig:energy}. Shortly after, the lattice is destroyed. Interestingly, during all this process (which lasts for $\approx 300 \tau_f$) the vorticity $\omega_\parallel$ remains approximately the same. The time scale of this process is orders of magnitude longer than the turnover time at the scale of these vortices. The evolution suggests that circulation and kinetic energy in each vortex keeps increasing  until viscous effects become relevant. However, while the amplitudes change in time, the radius of the vortex (associated with the lattice periodicity) does not change significantly.

Using the vortex centers we can perform a Voronöi tesselation \cite{Aurenhammer_1991}. The tesselation assigns to each vortex a cell, such that all points in that cell are closer to that vortex than to any other vortex (see Fig.~\ref{fig:voro}). The cells form polygons. From these cells we can count the number of vortices $N_V$ as the number of cells, compute the distance between vortices as the distance between pairs of centers of Voronöi cells $l$, and compute its mean value $\langle l \rangle$ and its dispersion $\sigma_l$ (which measures how regular the lattice is). We can also compute the lengths of the sides of the cells $l_C$, and their average $\langle l_C \rangle$, as well as the internal angles of the polygons $\theta_C$. For a perfectly regular lattice, we expect these quantities to be constant and show no dispersion. In fact, for a triangular lattice, all the Voronoï cells should be perfectly hexagonal.

With $l_C$ and $\theta_C$ we can define a second measure of lattice regularity. To this end we use the cell regularity measure \cite{chalmeta}, defined as
\begin{equation}
\mu = 1 - \frac{d(P,Q)}{1-\left(4-8/n\right)\pi},
\end{equation}
where $n$ is the number of vertices in the cell, and $d(P,Q)$ is the $L_1$ norm distance between two vectors $P$ and $Q$. The vector $P$ has $2n$ values, the first $n$ coordinates being the lengths of all cell sides $l_{C}$ normalized by the total perimeter of the cell, and the last $n$ being the internal angles of the cell $\theta_C$. The vector $Q$ is constructed with the first $n$ elements having the median of $l_{C}$, and the last $n$ elements equal to the internal angle of an $n$-sided regular polygon. In practice we use as a measure of lattice regularity the mean over all cells, $\langle \mu \rangle$.

Figure \ref{fig:stats} shows the time evolution of these quantites in simulation VII (the behavior shown in this figure is typical for all the simulations in table \ref{tab:sims}). At early times, $\langle l \rangle$ and $\langle l_C \rangle$ grow as the inverse cascade develops, but after $t \approx 500 \tau_f$ these quantities remain approximately constant for $\approx 700 \tau_f$. The dispersion $\sigma_l$ decreases and reaches a minimum in this period, while $\langle \mu \rangle$ reaches a maximum. Meanwhile, the number of vortices drops from $N_V \approx 30$ to 7, and remains constant for the same period of time. This is the time in which the lattice can be identified. In fact, the times at which $\sigma_l$ is minimum and $\mu$ is maximum are the times at which the lattice is more regular. Finally, after $t \approx 1200 \tau_f $ the lattice becomes unstable and $N_V$ decreases rapidly again.

Several quantities in Fig.~\ref{fig:stats} display a strong correlation: $\langle l \rangle$ and $\langle l_C \rangle$ are correlated, while $\sigma_l$ and $\langle \mu \rangle$ are anti-correlated. This is the result of certain properties of Voronöi tesselates. Given a tesselate on a plane with a well defined mean number of points per unit area, the probability of finding a certain amount of points on a fixed area is given by a Poisson distribution \cite{meijering, hinde}. It can then be shown that both $\langle l\rangle$ and $\langle l_C\rangle$ are only a function of the number of vortices per unit area $n_V = N_V/A_\perp$, such that
\begin{align}
    \langle l \rangle &= \frac{32}{9\pi} n_V^{-1/2}, & \text{and} & & \langle l_C \rangle &= \frac{2}{3} n_V^{-1/2}.
    \label{eq:voronoi}
\end{align}
Indeed, the validity of the relation between $\langle l \rangle$ and $n_V$ is explicitly verified in Fig.~\ref{fig:stats}.

We can thus identify vortex crystals in the simulations, and quantify its regularity, using $N_V$ and $\langle \mu \rangle$. Based on these quantities we define the ``crystal time'' $t_C$ as the period of time over which the number of vortices remains constant (without flucuations), and with a mean lattice regularity $\langle \mu \rangle$ above $0.75$. For simulation VII, this time is indicated by a double-headed arrow in Fig.~\ref{fig:stats}.

\subsection{Effect of the domain horizontal geometry \label{sec:geometry}}

The lattice could be an artifact of the domain geometry, with some resonance taking place at a particular scale given by $k_f$ and the smallest available wave number. Or its instability could be controlled by defects in the lattice, as a perfectly triangular lattice cannot be fitted in a square periodic domain. To confirm or discard these hypothesis we performed simulations with $\lambda_\perp \neq 1$. In particular, simulations IX, X, and XI in table \ref{tab:sims} have horizontal aspect ratios compatible with the translational symmetries of triangular lattices, and as a result should display less defects in their  crystals. All simulations with non-square horizontal sections develop crystals, and have values of lattice regularity $\langle \mu \rangle$ similar to the simulations with square horizontal sections. Moreover, when maintaining all control parameters fixed except for $\lambda_\perp$, the number of vortices in the domain changes, but the vortex density $n_V$ remains the same. Finally, the crystal time $t_C$ is not affected by changes in the domain horizontal geometry.

\subsection{Lattice lifetime}

As discussed in Secs.~\ref{sec:spatiotemp} and \ref{sec:geometry}, the lattice lifetime (i.e., the crystal time) is much larger than the turnover time of the vortices in the crystal, and is independent of the domain horizontal geometry. Moreover, the evolution of the energy, dissipation rate, and spectra in Figs.~\ref{fig:energy} and \ref{fig:spectrogram} indicate that the end of the lattice is followed by a fraction of its energy being dissipated, while the evolution of individual vortices in Fig.~\ref{fig:evolution} indicates that the circulation and the kinetic energy of each vortex grow until dissipation becomes relevant. Considering that the lattice has wave number $k^* = 2\pi/\ell^*$, and that $\ell^*$ can be estimated using Eq.~(\ref{eq:voronoi}) as $\ell^*\sim \langle l \rangle \sim n_V^{-1/2}$, we can estimate the dissipation time scale of the crystals as
\begin{equation}
    t_\nu \sim \frac{1}{\nu {k^*}^2} = \frac{{\ell^*}^2}{4\pi^2 \nu} \sim \frac{1}{4\pi^2\nu n_V}.
\end{equation}

\begin{figure}
    \centering
    \includegraphics[width=0.45\linewidth]{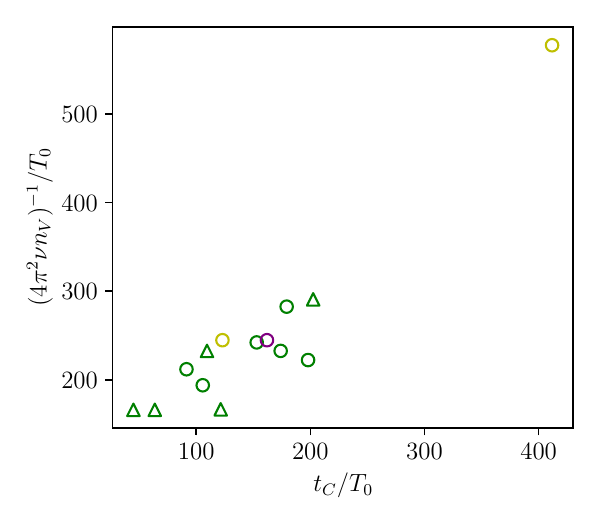}
    \caption{Dissipation time scale $t_\nu$ of a structure with periodicity $k^* = 2\pi/\ell^* \sim 2\pi n_V^{1/2}$ versus the crystal time $t_C$, for all simulations in table \ref{tab:sims} that develop a vortex crystal. Colors indicate the perpendicular aspect ratio with green for $\lambda_\perp=1$, purple for $\lambda_\perp=\sqrt{3}$, and yellow for $\lambda_\perp=\sqrt{3}/2$. Shapes of the symbols indicate the parallel aspect ratio of the simulations, with circles for $\lambda_\parallel=1$ and triangles for $\lambda_\parallel=2$.}
    \label{fig:tvisc}
\end{figure}

Figure \ref{fig:tvisc} shows the viscous time $t_\nu $ as a function of the crystal time $t_C$, for all simulations in table \ref{tab:sims} that develop a vortex crystal. 
Colors indicate the perpendicular aspect ratio with green for $\lambda_\perp=1$, purple for $\lambda_\perp=\sqrt{3}$, and yellow for $\lambda_\perp=\sqrt{3}/2$. The shapes of the symbols indicate the parallel aspect ratio of the simulations, with circles for $\lambda_\parallel=1$ and triangles for $\lambda_\parallel=2$. Irrespectively of the aspect ratio of the domains, and of other controlling parameters, the data are compatible with a linear relation $t_C \sim t_\nu$. This suggests that the mechanism that breaks the flux-loop energy transfer and destroys the lattice is the sustained growth of energy in the vortices, until dissipation at the lattice scale becomes dominant.

\subsection{Lattice vortex density and characteristic scale \label{sec:control}}

\begin{figure}
    \centering
    \includegraphics[width=.9\linewidth]{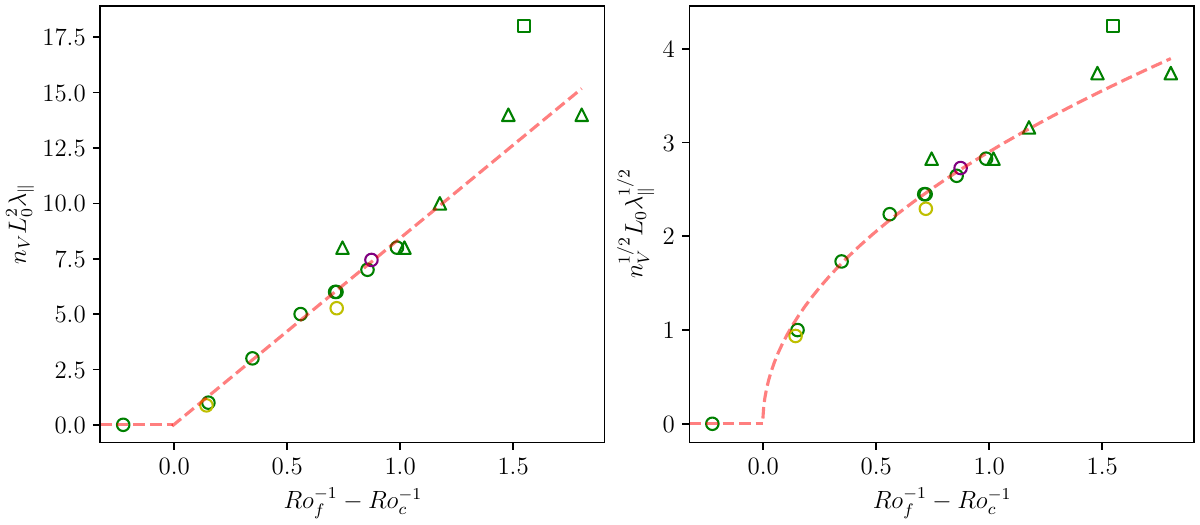}
    \caption{{\it Left:} Total number of vortices in a ``unit'' lattice, $n_v L_0^2$, times $\lambda_\parallel$, as a function of the inverse Rossby number $\textrm{Ro}_f^{-1}$ displaced by a critical Rossby number.  {\it Right:} $n_V^{1/2}$ (proportional to the lattice wave number $k^*$) normalized as in the left panel, as a function of the inverse Rossby number and displaced by the critical Rossby number. Labels for all circles and triangles are as in Fig.~\ref{fig:tvisc}. The green square corresponds to a simulation using hyperviscosity from \cite{clark}. The red dashed lines in both panels correspond to the relations given by Eq.~(\ref{eq:scaling}).
    }
    \label{fig:scaling}
\end{figure}

To this point, it is still unclear what parameters set the lattice periodicity. In \cite{clark} it was shown, using simulations with hyperviscosity, that there is a critical Rossby number for the lattice to appear, and a second critical Rossby number for a pure inverse cascade without metastable states to develop. For $\textrm{Ro}_f^{-1}$ below the inverse of the first critical Rossby number, rotation is too weak and the flow displays only a direct energy cascade. For $\textrm{Ro}_f^{-1}$ above the inverse of the second critical Rossby number, rotation is too strong and energy piles up only at the largest available scale. We will define the critical Rossby number $\textrm{Ro}_c$ as the first critical value, needed to obtain vortex crystals in a flow at given domain aspect ratio. Data in \cite{clark} indicate that the development of a vortex crystal depends on $\lambda_\parallel$, while our simulations discussed in Sec.~\ref{sec:geometry} indicate that the lattice formation and the properties of the crystal do not depend on $\lambda_\perp$. Moreover, the data in \cite{clark} suggests that $\textrm{Ro}_c$ varies as $\lambda_\parallel^{1/2}$. We can thus define a rescaled critical Rossby number for all simulations with different vertical aspect ratios in table \ref{tab:sims} as
\begin{equation}
\mathrm{Ro}_c = \mathrm{Ro}_c^* \lambda_\parallel^{1/2},
\end{equation}
where we set $\mathrm{Ro}_c^*$ as the critical value to obtain a crystal when $\lambda_\parallel=1$.

The viscosity (or the Reynolds number) affects the lifetime of the crystal, but does not seem to affect the number of vortices in the lattice. Changing $\lambda_\perp$ changes the number of vortices as the total horizontal area may change, but the vortex density remains the same. Exploration of table \ref{tab:sims} indicates that the number of vortices grows approximately linearly with $\Omega$ in simulations in which the other parameters remain fixed. Finally, careful exploration of the table and of the crystals in simulations XIV, XV, and XVI, with $\lambda_\parallel = 2$, indicate that beside changing $\textrm{Ro}^*_c$, $\lambda_\parallel$ also affects the number of vortices and the crystal periodicity. 

We thus propose
\begin{equation}
    \widetilde{n}_V = n_V L_0^2 = \frac{1}{\lambda_\parallel} \left( \frac{1}{\mathrm{Ro}_f} - \frac{1}{\mathrm{Ro}_c} \right) ,
    \label{eq:scaling}
\end{equation}
and $\widetilde{n}_V = 0$ when $\mathrm{Ro}_f^{-1} < \mathrm{Ro}_c^{-1}$, where $\widetilde{n}_V$ is the dimensionless vortex density (which can be also interpreted as the number of vortices in a crystal in a domain with unit area). The left panel of Fig.~\ref{fig:scaling} shows all simulations in table \ref{tab:sims}. The data shows good agrement with Eq.~(\ref{eq:scaling}), even when including one simulation using hyperviscosity from \cite{clark} with very different parameters, marked by a square. We can finally estimate the characteristic length or typical wave number of the crystal. Using the relation $k^* \sim n_V^{1/2}$, we expect $k^* \sim \lambda_\parallel^{-1/2}  (\mathrm{Ro}_f^{-1} - \mathrm{Ro}_c^{-1})^{1/2}$. The data from the DNSs in table \ref{tab:sims} is compared against this expression in the right panel of Fig.~\ref{fig:scaling}, showing good agreement.

\section{Conclusions}

We used DNSs of the incompressible Navier-Stokes equations in a rotating frame to study vortex crystal formation in classical turbulent flows. The simulations were conducted in triply periodic domains with varying horizontal and vertical aspect ratios. Energy was injected at intermediate scales, using normal viscosity which resulted in moderate Reynolds numbers. 
The primary control parameters were the Rossby number $\textrm{Ro}_f$ at the forcing scale, the Reynolds number $\textrm{Re}_f$ at the forcing scale, and the domain geometry. By analyzing energy spectra, vortex morphology, and spatio-temporal flow properties, we identified cases in which crystals form in metastable states and their lifetimes. The results showed that vortex crystals form transiently as a result of energy stalling at intermediate scales, caused by a flux-loop mechanism \cite{clark}. The crystal’s lifespan is linearly proportional to the viscous time at the crystal scale. The periodicity of the vortex lattice depends on $\textrm{Ro}_f$ and on the vertical domain aspect ratio $\lambda_\parallel$, with a critical Rossby number $\textrm{Ro}_c$ marking the transition to crystal formation. The findings indicate dynamics reminiscent of phase transitions, with implications for self-organization in rotating turbulent flows, connecting the formation of vortex crystals with other critical phenomena observed in flows with inverse energy cascades \cite{biferale, ABB2018, clark}.

The scaling laws identified in this work have other implications. The dependence of the wave number $k^*$ with the domain vertical aspect ratio $\lambda_\parallel$ suggests that lattice formation is controlled by the number of near resonant interactions available in the system. Note that the density of modes in Fourier space in the vicinity of $k_z \approx 0$ is controlled by $\lambda_\parallel$, which in turn controls the number of resonant and near resonant triads \cite{Clark_di_Leoni_2016b}. On the one hand, for $\lambda_\parallel \to \infty$ there is a continuous distribution of vertical wave numbers $k_z$. In this limit, for a fixed Rossby number, the density of large-scale vortices goes to zero, which corresponds to the regime in which the two-dimensional modes decouple from the three-dimensional modes, and no inverse cascade develops \cite{cambon, Cambon_2004}. On the other hand, for too small $\lambda_\parallel$, the density of modes in Fourier space in the vicinity of $k_z \approx 0$ is very small. In this case $k^*$ can become larger than $k_f$, and no lattice can form. This also explains why in \cite{clark} two critical Rossby numbers were found, one for the lattice to form, and a second critical value for inverse cascades to develop without lattices. At fixed $\lambda_\parallel$, as $\textrm{Ro}_f$ decreases, eventually $k^*$ becomes larger than $k_f$ and no lattice can form at scales larger than the forcing scale.

These results mean that in rotating flows, not too shallow, not too deep domains are needed for vortex crystals to develop. In planetary atmospheres, this would imply that ordered arrays of vortices require deep atmospheres, and that once formed these structures can last for very long times (of the order of their viscous time). However, the results presented here have several shortcomings that must be considered. First, only the effect of uniform rotation in homogeneous flows is considered in this study. In many vortex arrays observed in nature, boundary conditions, planet curvature, stratification, and convection play relevant roles. Moreover, the simulations presented here have relatively modest Reynolds numbers, as normal viscosity is used. While this choice allowed us to identify the physical time scale associated with the lifetime of the lattice, it results in a limitation in our ability to explore parameter ranges relevant for experiments or for observations.

\begin{acknowledgments}
The authors acknowledge support from UBACyT Grant No.~20020220300122BA, and from project REMATE of Redes Federales de Alto Impacto, Argentina. 
\end{acknowledgments}

\bibliography{ms}

\end{document}